\documentclass[12pt]{article}
\usepackage{verbatim}

\title{Gauge Invariance and Symmetry Breaking by Topology and Energy Gap}

\author{
Carlo Heissenberg\footnote{e-mail: carlo.heissenberg@sns.it} \ and F. Strocchi\footnote{e-mail: franco.strocchi@sns.it} \\ Scuola Normale Superiore, Pisa, Italy}

\date{}

\makeglossary

\newtheorem{Theorem}{Theorem}[section]

\newtheorem{Proposition}[Theorem]{Proposition}

\usepackage{latexsym}
\usepackage[T1]{fontenc}
\usepackage{amsmath}
\usepackage{amssymb}
\usepackage{hyperref}
\hypersetup{
    colorlinks=true,
    linkcolor=blue,
    filecolor=magenta,      
    urlcolor=cyan,
    citecolor=magenta,
}

\def \AO {{\cal A}({\cal O})}
\def \AO' {{\cal A}({\cal O}')}

\def \at {{\alpha_t}}

\def \Pf {{\bf Proof.\,\,}}

\def \be {\begin{equation}}
\def \ee {\end{equation}}

\def \ra {\rightarrow}

\def \eqq {\equiv}

\def \a {{\alpha}}
\def \b {{\beta}}
\def \g {{\gamma}}
\def \d {{\delta}}

\def \l {{\lambda}}

\def \s {{\sigma}}

\def \t {{\tau}}

\def \om {{\omega}}

\def \A {{\cal A}}

\def \F {{\cal F}}
\def \G {{\cal G}}
\def \H {\mbox{${\cal H}$}}

\def \K {{\cal K}}

\def \M {{\cal M}}

\def \Z {{\cal Z}}

\def \id {{\bf 1 }}

\def \Psio {{\Psi_0}}

\def \d^nu {{\partial^\nu}}
\def \d^la {{\partial^\lambda}}
\def \d^o {{\partial^0}}

\def \Cbf {{\bf C}}

\def \Real {\mathbb{R}}
\def \Zint {\mathbb{Z}}

\def\doppio#1{{\rm I}\kern-.1667em{\rm #1}}

\def\Q{\text{Q}\kern-.52em
    \text{\vrule height1.5ex width.5pt depth0pt}\kern.45em}

\def\dZ{{\mathchoice {\hbox{$\Ss\textstyle Z\kern-0.4em Z$}}
{\hbox{$\Ss\textstyle Z\kern-0.4em Z$}} {\hbox{$\Ss\scriptstyle
Z\kern-0.25em Z$}} {\hbox{$\Ss\scriptscriptstyle Z\kern-0.2em
Z$}}}}

\def\dC{{\mathchoice{\hbox{$\rm\textstyle\text{\kern.35em\vrule
   height1.5ex width.5pt depth0pt\kern-.35em C}$}}
{\hbox{$\rm\textstyle\text{\kern.35em\vrule
   height1.5ex width.5pt depth0pt\kern-.35em C}$}}
{\hbox{$\rm\scriptstyle\text{\kern.35em\vrule
   height1.5ex width.3pt depth0pt\kern-.35em C}$}}
{\hbox{$\rm\scriptscriptstyle\text{\kern.35em\vrule
   height1.5ex width.2pt depth0pt\kern-.35em C}$}}}}

\makeatletter

\@addtoreset{equation}{section}

\begin{document}

\maketitle
\begin{abstract}
For the description of observables and states of a quantum system, it may be convenient to use a canonical Weyl algebra of which only a subalgebra $\A$, with a non-trivial center $\Z$,  describes observables, the other Weyl operators playing the role of intertwiners between inequivalent representations of $\A$.  In particular, this gives rise to a gauge symmetry described by the action of $\Z$. A distinguished case is when the center of the observables arises from the  fundamental group of the manifold of the positions of the quantum system. Symmetries which do not commute with the topological invariants represented by elements of $\mathcal Z$ are then spontaneously broken in each irreducible representation of the observable algebra, compatibly with an energy gap; such a breaking exhibits a mechanism radically different from Goldstone and Higgs mechanisms.  
This is clearly displayed by the quantum particle on a circle, the Bloch electron and the two body problem.        
\end{abstract}

\newpage
\section{Introduction}
   The mathematical foundations of Quantum Mechanics rely on the Dirac-von Neumann axioms (for a critical review see \cite{FS1,FS2}) and the equivalence between the Heisenberg formulation in terms of canonical operators  and the Schr\"odinger formulation in terms of wave functions is provided by the Stone-von Neumann theorem which states the uniqueness of the Schr\"{o}dinger representation of the canonical algebra under general regularity conditions.

Technically, this result is achieved  by introducing the Weyl unitary operators (formally the exponentials of the canonical  variables $q, p$) and the corresponding Weyl algebra, defined by algebraic relations which encode the canonical commutations relations of Heisenberg canonical variables.  
 
\def \d {{\delta}}
The use of the Weyl algebra 
is usually motivated  by the better behavior and
mathematical control of the unitary Weyl operators with respect to the
Heisenberg canonical variables, which are necessarily represented by unbounded 
operators. 

It is usually taken for granted that the Dirac-Heisenberg quantization, in terms of the canonical commutation relations of the $q$'s and $p$'s, and the Weyl quantization in terms of the commutation relations of the Weyl operators are  equivalent under the implicit assumption that one is interested only in regular representations of the Weyl algebra, to which the Stone-von Neumann uniqueness theorem applies.

The point is that 
the Dirac-Heisenberg 
quantization implicitly assumes that all of the canonical variables describe 
observables so that the regularity of their exponentials (Weyl operators) 
is required by their existence as (unbounded) operators in the Hilbert 
representation space.

The regularity condition at the basis of the Stone-von
Neumann theorem
is standard in the mathematical analysis  and classification of Lie group
representations and in the quantum mechanical case it amounts to consider the
strongly (equivalently weakly) continuous (unitary) representations of the
Heisenberg group.

However, for a class of physical
systems, especially  in connection with quantum gauge
theories, it has become apparent  \cite{AMS1,AMS2,AMS3,FSS} that the Dirac-Heisenberg quantization 
is not compatible  with a gauge 
invariant ground state, only the
Weyl quantization being allowed.

In these cases, the inequivalence of the two
quantization methods arises by the lack of regularity of one-parameter
groups generated by  Weyl operators, so that the corresponding generators,
i.e. the  corresponding Heisenberg canonical  variables, cannot be defined as (self-adjoint)
operators in the Hilbert space of states.

The physical reason at the basis of such a {\em lack of regularity} is that the QM
description of a class of physical  systems involves
canonical variables, not all of which correspond to observable quantities; 
some of them are introduced for the description of the states, namely of the inequivalent
representations of the algebra of observables, for which purpose   only  their 
exponentials
are needed  to exist as  well defined operators in the Hilbert space of states, with the role of  {\em intertwiners between inequivalent representations}. 

Thus,   the larger  algebra $\F$, in which the algebra of observables is embedded, may be characterized as the minimal algebra such that all of the automorphisms of $\A$ are inner. This means that $\F$ contains the intertwiners between the inequivalent representations of $\A$, i.e. all of the ``variables'' needed for a complete description of the observables {\em and} the states of the given quantum system. Such an embedding provides a natural $C^*$ norm for $\A$. Clearly, the basis of such a structure, beyond the Stone-von Neumann unique characterization of quantum mechanics, is the existence of inequivalent representations of $\A$. This typically occurs in the case of quantum systems  on manifolds with a non-trivial fundamental group.

Thus, what might at first sight look as an uninteresting singular, if not
pathological, case turns out to be crucial for the quantum description of
physically interesting systems.

In general, this lack of regularity may be related to the existence of 
a gauge group. Typically, one has the Weyl algebra $\A_W$ generated by the 
exponentials of the full set of  canonical variables needed for the description 
of the observables {\em and} states of the system, but only a subalgebra $\A$ describes observables. 

Generically, $\A$ has a non-trivial center $\Z$, which generates transformations 
having the meaning of gauge symmetries ({\em  gauge
transformations}). Thus, the algebra $\A_W$ of canonical variables
contains both gauge dependent and  gauge invariant (i.e. observable) variables. 

Clearly,  the regularity condition must be satisfied by the exponentials
of the observable  variables, otherwise the representation is not 
physically acceptable,  but there is no physical reason for the regularity condition of the
gauge dependent Weyl operators, playing the role of intertwiners. 

As we shall see, the  representation of the Weyl
algebra by a gauge invariant ground state in general requires the non-regularity 
of the gauge dependent Weyl operators and implies the impossibility of defining the
corresponding generators as well defined operators in the corresponding Hilbert
space ({\em non-regular representation of the Heisenberg group or of the Weyl
algebra}).

Relevant quantum mechanical examples of such a structure are the electron in a
periodic potential ({\em Bloch electron}),  the {\em Quantum Hall electron},
the {\em particle on a circle}, where the gauge transformations are,
respectively, the lattice translations, the magnetic translations and the
rotations of $2 \pi$. In Gauge Quantum Field Theories (GQFT)  the need of a 
non-regular Weyl quantization arises in (positive) representations of the field 
algebra defined by a gauge invariant vacuum state.

The non-regular Weyl representations for the quantization
of  systems with a gauge symmetry exhibit the following
characteristic structures, which do not
have a counterpart  in the Dirac-Heisenberg canonical
quantization: \newline i) a {\em gauge invariance constraint} in operator form
 compatible with canonical Weyl
quantization (avoiding the mathematically unacceptable  recourse to 
non-normalizable  states)
  \newline ii) {\em superselected
charges} defined by the center of the observable algebra \newline
iii) {\em gauge invariant ground states}, defining inequivalent representations 
of the observable
algebra, labeled by the spectrum of the superselected charges ($\theta$ sectors)
\newline iv) {\em absence of ``Goldstone states''} associated  to the spontaneous 
breaking of symmetries (canonically)
conjugated to the  gauge transformations. 

Particularly relevant is the case in which the above structure of the algebra of observables for a (finite dimensional) quantum system arises as a consequence of the non-trivial topology of the manifold $\mathcal M$ of the particle positions. The mechanism is that the fundamental group of the manifold, which has been shown in \cite{MS} to be the only source of topological effects, gives rise to elements of the center of the observable algebra and a corresponding gauge group. 
This leads to a  mechanism of symmetry breaking which is substantially different  from the standard Nambu-Goldstone  case (with associated Goldstone bosons), and from the Higgs mechanism (characterized by a dynamics which induces a long range Coulomb like delocalization): the symmetry breaking is forced by the topology in any irreducible representation of the observables, with a compatible energy gap related to the spectrum of the first homology group of the manifold ({\bf symmetry breaking by topology and energy gap}, see Section 5 below).      

In our opinion, the realization of such  general structures and their
clear and simple realization in (finite dimensional) QM mechanical models,
fully under control, sheds light on the more difficult 
infinite dimensional
GQFT models, discussed within very specific approximations.

The physical relevance of non-regular representations of the Heisenberg group
raises the problem of the their classification, i.e.  a generalization of the
classical Stone-von-Neumann theorem \cite{Stone,vonNeumann,Mackey,Rieffel}, which characterizes the regular ones.

Such a generalization
may be obtained by exploiting the simple form of the Gelfand spectrum of 
the  maximal abelian subalgebra $\A_Z$ of the Weyl algebra generated by the pairs 
$U_i(- 2\pi/\lambda)$, $V_i(\l)$, $ i = 1, ...d$, ($d$ the space dimensions), formally  
corresponding to the exponentials 
$\exp{ -i (2\pi/\l) q_i}$, $\exp{i \l\, p_i}$.
Such pairs of operators were introduced by Zak \cite{ZakOr} in order to discuss the dynamics of electrons in solids in external fields; the crucial distinctive property (see \cite{CMS}) of the corresponding $C^\ast$-algebra, called the 
{\em Zak algebra}, is that its Gelfand spectrum $\Sigma$ is given by $d$ copies of the 
two-dimensional torus $ \Sigma = ({\bf T}^2)^d$. 

Then, one may prove \cite{CMS} that all  the representations 
of the Weyl algebra which are {\em spectrally multiplicity free} as representations 
of its Zak algebra (a condition which generalizes irreducibility) and are 
{\em strongly measurable} (a condition which replaces regularity in non-separable spa\-ces) 
are unitarily equivalent to a representation of the Weyl algebra of the same 
form of the standard Schr\"{o}dinger representation on $L^2(\Sigma, d\mu)$, 
with $d\mu$ a (positive) translationally invariant Borel measure, which reduces 
to the Lebesgue measure iff the regularity condition is satisfied. 

The conditions which yield such a classification are satisfied by all of the 
non-regular representations of physical interest mentioned above;
thus, they all have the above form, each with a corresponding Borel measure on 
$({\bf T}^2)^d$.
     
This represents a radical departure from the standard structure of Quantum Mechanics, since it requires non-separable Hilbert spaces, very discontinuous expectations of one-parameter groups of unitary operators (vanishing for all non-zero values of the parameters, so that the corresponding generators do not exist), etc. Such features have been regarded as pathological and to be avoided for an acceptable physical interpretation. Actually,  they have a very sound mathematical status, there is no problem for the physical interpretation and  they explain the important role of the gauge group and of topology for the evasion of Goldstone theorem in the case of gauge symmetry breaking.
\vspace{1mm}

The role of topology in (flat space) GQFT, especially in Quantum Chromodynamics (QCD),  has been realized a long time ago, but in terms of topological classification of the euclidean configurations (the instanton solution in QCD) assumed to dominate the functional integral within a semiclassical approximation. Such a classification requires regularity and continuity of such euclidean configurations and it is well known that regular configurations have zero functional measure. Apart from this consistency problem, the above exploitation of topology appears to be strictly bound to very special mechanisms and the question arises of whether there is an underlying general framework, a point not clearly addressed in literature. 

One the aims of this paper is to emphasize, on the basis of quantum mechanical models recognized to mimic the basic structures of GQFT, that the crucial and basic ingredient is the existence of a \emph{nontrivial center of the observables} and its \emph{topological origin}. This clarification, in terms of the topology of the gauge group, provides a general abstraction of the mechanism, similar to that provided by studying the properties of a Lie group rather than those of the vectors of one of its representations (Section 2).

Such a topological origin of the center of the observable algebra arises as a consequence of the topology of the manifold of the QM configurations (Section 3 and examples, which are reviewed under this more general perspective).

The main focus of the paper is to study the implications on spontaneous \emph{breaking of symmetries not commuting with the topological invariants} which define elements of the center $\mathcal Z$ of the observables. The conclusion (Theorem 5.2) is that such symmetries are broken in any irreducible (or factorial) representation of the observables and that (contrary to the classical Goldstone theorem) the energy spectrum may have a gap given by the spectrum of the topological invariants belonging to $\mathcal Z$.


\section{Gauge invariance, superselection rules  \\and non-regular canonical quantization} 

\def \GO {G_{obs}}
A general case  leading to non-regular representations is when 

\noindent i) a 
quantum system is described by canonical
variables, generating a Heisenberg group $G_H$,
but only a subset of them, and consequently only a subgroup $G_{obs} \subset
G_H$,
 describes observable
quantities, the other canonical variables providing the intertwiners for the description of the inequivalent representations of the observable algebra  

\noindent ii) $\GO$ is generated
by a Heisenberg subgroup and by an abelian subgroup $\G$ which commutes with $\GO$. 

Then,  $\G$ generates a group of
transformations $\a_g$, $ g \in \G$, which leave the observables pointwise
invariant \be{\label{agAA}\a_g(A) = A, \,\,\,\,\,\forall A \in \GO,\,\,\,\,\forall g \in
\G,}\ee i.e. $\G$ has the 
meaning of a {\bf gauge group},

The elements of $G_H$ generate a
$C^*$-algebra $\F_W$,  called {\em   field algebra}, and the elements
of $\GO$ generate a {\em $C^*$-algebra $\A$ of
observables},
characterized by  gauge invariance,
eq.\,\eqref{agAA}. 
$\A$ has a non-trivial {\em center}
 generated by
the elements of $\G$. A representation of $\F_W$ is {\em
physical}
if $G_{obs}$ is regularly
represented.

In the irreducible representations  of $\A$,  $\Z$ is represented by  multiples 
of the identity. The generators of $\G$
have the meaning of {\bf superselected charges}
and the points $\theta$ of the spectrum $\sigma(\Z)$ of $\Z$ label inequivalent 
representations
$(\H_\theta, \pi_\theta)$ of $\A $, called \mbox{\boldmath$\theta$}
{\bf sectors}.
 Stone-von Neumann
uniqueness theorem does not apply and this can be traced back to the fact that,
contrary to the Weyl $C^*$-algebra, $\A$ is not simple.

By definition a {\em gauge invariant state} $\om$
on $\F_W$ satisfies $$\om(\a_g(F)) = \om (F), \,\,\,\,\forall F \in \F_W$$ and
therefore, in the GNS representation $\pi_\om$ of $\F_W$ defined by $\om$, the
gauge transformations are implemented by unitary operators $U(g)$ defined by 
($\Psi_\om$ denotes the vector which represents $\om$) $$U(g) \Psi_\om =
\Psi_\om,\,\,\,\,U(g) \pi_\om(F) \Psi_\om = \pi_\om(\a_g(F))\,\Psi_\om,
\,\,\,\,\forall F \in \F_W.$$

Let  $V(g)$ denote the element of $\G$ which defines $\a_g$:  $$\a_g(F) = V(g)\,
F\, V(g)^{-1}, \,\,\,\,\forall F \in \F_W;$$ then, $\pi_\om(V(g)) U(g)^* $ commutes
with $\F_W$ and, in each irreducible representation of $\F_W$,
$\pi_\om(V(g))\,U(g)^*
 = e^{i \theta(g)} \id$. Hence,   $\Psi_\om$ is an eigenvector of
$\pi_\om(V(g))$, with eigenvalue $e^{i \theta(g)}$,   \be{\pi_\om(V(g)) \Psi_\om 
= \pi_\om(V(g)) U(g)^* \Psi_\om =
e^{i \theta(g)} \Psi_\om.}\label{eitheta}\ee

\def \psiom {\Psi_\om}
Thus, the GNS
representation $\pi_\omega$ of $\F_W$, equivalently of $G_H$, defined by a 
gauge invariant
state $\omega$ is non-regular, $$\H_{\pi_\omega} =
\bigoplus_{\theta \in \s(\Z)}\H_{\theta},$$ and the subspaces $\H_\theta$
carrying disjoint irreducible representations of $\A$ are proper
subspaces of the non-separable space $\H_{\pi_\omega}$.\\

Summarizing, one has:
\begin{Proposition} Let $G_H$ be the Heisenberg group defined by the set of 
canonical variables
 $\{q_i, \,p_i\}$,   $\F_W$  the corresponding canonical $C^*$-algebra,  
$\A \subset \F_W$
 the $C^*$-subalgebra of observables
and $\G$ the commutative group of gauge transformations, defined by a subgroup
$\G \subset G_H$.

Then, the GNS representation of $\F_W$ defined by a gauge invariant state is a
non-regular representation of  $\F_W$, as well as  of the Heisenberg group
$G_H$, and the elements of $\G$ define {\bf superselection
rules}.
\end{Proposition}

 Relevant examples of such a structure are provided by quantum
mechanical models, in particular those exhibiting strong analogies with gauge
quantum field theories (as discussed below).

The prototype is provided by the non regular representation \cite{BMPS} of Weyl algebra $\A_W(\Real^2)$ generated  by the one-parameter unitary groups 
$U(\a), \,V(\b)$, $\a, \b \in \Real$, with  $U(\a) V(\b)  = V(\b) U(\a)
e^{- i \a \b}$, $V(\b)$ describing gauge transformations.  

\begin{Proposition} The GNS representation $(\pi_0,  \H_0)$ of  $\A_W(\Real^2)$ defined by a pure
gauge
invariant state $\om_0$ is unitarily
equivalent to the following representation \be{\omega_0(U(\a) \,V(\b)) = 0,
\,\,\,\,\mbox{if} \,\,\,\,\a \neq 0, \,\,\,\,\,\omega_0(V(\b)) = e^{i \b
\bar{p}}, \,\,\,\bar{p} \in \Real.}\label{omega0UV}\ee
Thus, the one-parameter group $U(\a)$ is non-regularly represented. The GNS
representation space $\H_0$ contains as representative of $\om_0$ a cyclic
vector $\Psio$ such that (denoting by the same symbols the elements of the Weyl
algebra and their representatives) \be{ V(\b) \Psio = e^{i \b
\bar{p}}\,\Psio,\,\,\, \,\,\,\,(U(\a) \,\Psio, \,U(\a')\,\Psio) = 0,
\,\,\,\,\mbox{if}\,\,\,\a \neq \a' .}\ee The linear span $D$ of the vectors
$U(\a) \Psio$, $ \a \in \Real$ is dense in $\H_0$, which is therefore  non-separable.

The generator of the  one-parameter group $U(\a)$ does not exist, but
nevertheless a generic vector of $D$ $$\Psi_A = A \Psio, \,\,\,\,A = \sum_{n
\in \Zint} a_n U(\a_n), \,\,\,\,\{a_n\} \in l^2,$$ may be represented by a wave
function $\psi_A(x)=\sum_{n \in \Zint} a_n\, e^{i \a_n x}$, with scalar product
given by the ergodic mean \be{(\psi_A, \psi_A) = \sum_{n \in \Zint} |a_n|^2 =
\lim_{L \ra \infty} (2 L )^{-1} \int_{-L}^L d x
\,\bar{\psi}_A(x)\,\psi_A(x).}\ee

The spectrum of $V(\b)$ is a pure point spectrum.
\end{Proposition}

\Pf\, By the gauge invariance of $\omega_0$ and the Weyl relations one has, $\forall \gamma \in\mathbb{R}$,
\begin{align*}
\omega_0(U(\alpha)V(\beta)) &= \omega_0(V(\gamma)U(\alpha)V(\beta)V(\gamma)^\ast)=\\
&=\omega_0(U(\alpha)V(\gamma)V(\beta)V(\gamma)^\ast) e^{i\alpha\gamma}=\omega_0(U(\alpha)V(\beta)) e^{i\alpha\gamma},
\end{align*}
which proves the first of eq.\eqref{omega0UV};
furthermore by eq.\eqref{eitheta} and the group law
$$
V(\beta)\Psi_0 = e^{i \bar{p}\beta}\Psi_0,\ \bar{p} \in \Real.
$$ 
The rest of the proposition easily follows.\\

\noindent {\bf Examples}
 
\noindent
1. {\em Gauge invariance in the two-body problem}

\noindent
The quantum description of two interacting particles is given by the Weyl algebra $\A_W$, corresponding  to the two particle canonical variables $q_1, q_2, p_1, p_2 $; however,   
for  the  discussion of  the bound state
spectrum of the two-body problem and in particular the lowest energy level, the position of the  center
of mass is irrelevant. 

It is therefore natural to consider as observable
$C^*$-algebra $\A$ the algebra generated by the relative canonical variable $q, p$ and by the center of mass momentum $P$.

Hence, the translations $v(\b)$ of
the center of mass have the meaning of {\em gauge transformations}.
The lowest energy state $\om_0$ must
satisfy $\om_0(P^2) = 0$, so that the corresponding vector $\Psio$ satisfies
$P^2 \Psio = 0$, i.e. it  is gauge invariant $v(\b) \Psio = \Psio$. 
This condition is incompatible with the canonical
commutation relations in the Heisenberg form.
It has been suggested to bypass such
incompatibility by allowing $\Psio$ to be non-normalizable. \cite{J}

 In our opinion,
such a choice would have catastrophic consequence on the GNS representation
defined by such a ground state; by the cyclicity of $\Psio$ all vectors of such
a representation would be non-normalizable,  all matrix elements (including the
ground state expectations of gauge invariant operators) would be divergent and
one could not extract finite results in a consistent mathematical way.

A canonical quantization is not forbidden, provided it is done
in terms of the Weyl algebra, rather than of the Heisenberg algebra, and it is given by Proposition 2.2

In our opinion, from a mathematical point of view, the non-re\-gu\-la\-ri\-ty of the
representation is a much better price to pay, rather than living with non-normalizable 
state vectors. 

The advantages of such a quantization is that the
states are described by {\em normalizable} vectors of a Hilbert space,  the
basic quantum mechanical rules are not violated, the observable subalgebra $\A$
is regularly represented in the standard way,  the canonical variables which
are not gauge invariant are non-regularly represented,  only their exponentials
being well defined.
 
Thus, the model suggests a general strategy for quantizing systems with a gauge invariance constraint. 

\vspace{2mm}
\noindent 2. {\em Bloch electron}

Another relevant quantum mechanical example is provided by an electron in a periodic bounded measurable potential $W(q) = W(q + a)$. The periodic
structure of the system leads to  consider as observable $C^*$-algebra $\A$ the subalgebra of the Weyl algebra $\F_W$,
generated by the translations $V(\b)$ and by the periodic functions of the position $U(2
\pi n/a)$, $n \in \Zint$.

The center $\Z$ of $\A$ is generated by the translations $V(a)$ and the irreducible
representations of $\A$ are given by the subspaces $\H_\theta$ ({\em $\theta$
sectors}), corresponding to the GNS representations of $\A$ defined by the states invariant under the gauge group of translations $V(a)$. 

The operators $U(\a/a), \a \neq 2 \pi n$ do not commute with the center of $\mathcal A$ and therefore intertwine between  the
inequivalent representations $\pi_\theta$ and $\pi_{\theta + \a}$; the
corresponding one-parameter group is non-regularly represented in  the
representation of $\F_W$  defined by the gauge invariant ground state.\cite{LMS}
 

\section{Topology,  gauge groups and Weyl non-regular quantization} 

The structure of an observable algebra $\A$  naturally embedded in a larger Weyl algebra 
$\F_W$, with the non-trivial center of $\A$ generating gauge transformations on 
$\F_W$, arises for example as a consequence of the non-trivial topology of the manifold $\M$ which 
describes the configurations of the quantum system. 

In fact, the quotient of the fundamental group $\pi_1(\tilde{\M)}$, of the universal covering space  $\tilde{\M}$ of the manifold, with its commutator group, i.e. the first  homology group $H_1(\tilde{\M)}$ of $\tilde{\M}$,  defines topological invariants which are represented by elements of the center of the observable algebra \cite{MS}. 

\vspace{1mm}
For simplicity, we consider the case of a {\em quantum particle on a circle}, where 
the observable algebra $\A$ can be taken as the $C^*-subalgebra$ of the standard Weyl (field) algebra $\F_W$,  generated by 
$U(n) = e^{ i n \varphi}$, $ n \in \Zint$, and $V(\b) = e^{ i \b p}$, $ \b \in
\Real$; with the canonical commutation relations  \be{ U(n)\, V(\b) = e^{- i n \b}
\,V(\b)\,U(n).}\ee

The rotations of $2 \pi$ define elements of the observable    
algebra $\A$, which may 
actually be characterized as the subalgebra of $\F_W$ invariant under
the translations $\g^n$ of $2 \pi n$,  $n \in \Zint$,  which, therefore, get the
meaning of {\em gauge transformations}.

The structure  fits into the general
discussion of Heisenberg group $G_H$, observable subgroup $G_{obs}$ and gauge
group  $\G$, with corresponding $C^*$-algebras $\F_W$, $\A$ and a non-trivial
center $\Z$ of $\A$, as discussed above. The {\em non-trivial center of $\A$ may be traced back to the non-trivial fundamental group of the circle}: $\pi_1(S^1) = \Zint$. The representation $\pi $ of
$\A$ is regular if $\pi(V(\b))$ is a weakly continuous group of unitary
operators.

In each irreducible representation of $\A$, the element $V(2\pi) $ is a
multiple of the identity, say $e^{i \theta}$, $ \theta \in [0, \,2 \pi)$.
\begin{Theorem} \rm{\cite{FS1}} For any given $\theta$, all of the irreducible regular
representations $\pi_\theta$ of $\A$ with $\pi_\theta(e^{i 2 \pi p }) = e ^{i
\theta}$ are unitary equivalent.

The Hilbert space $\H$ of the unique regular representation $\pi_S$ of the Weyl algebra $\F_W$    decomposes 
as a direct integral 
$$\H= \bigoplus_{\theta\in[0,{2\pi})}\H_\theta,$$
over the spectrum of $\pi_S(V(2 \pi)$. There is a unique irreducible representation $\pi$ of $\F_W$, whose Hilbert space decomposes as a direct sum  of the irreducible representations of the observable algebra $\A$; in such a representation $V(\b)$ is regularly represented, but the algebra generated by  $U(\a)$, $\a \in \Real$ is not. The operators $U(\a)$ intertwine between inequivalent representations  of $\A$.
\end{Theorem}

\vspace{1mm}
A similar structure is displayed by an {\em electron  in a periodic crystal, i.e. subject to 
 a periodic potential}.  For simplicity we
consider the one-dimensional case. In this case the  Hamiltonian  is $
H = - d^2/d x^2 + W(x)$, with the potential satisfying the periodicity
condition $W(x + a) = W(x)$, for a suitable $a$.
The field algebra $\F_W$ is generated by the Weyl operators $U(\a), V(\b)$, $\a
, \b \in \Real$. 

As in the case of a particle on a circle, the center of the observable algebra may be viewed as arising from the non-trivial topology of $\Real/ [0, a]$. 
In this case, the non-regular representations of $\F_W$ are defined by the gauge invariant ground states 
($\theta$ vacua).
\begin{Proposition} \rm{\cite{LMS}}
Let $W(x)$ be a bounded
measurable periodic potential, $W(x) = W(x + a)$, then there exists one and
only one irreducible representation $(\pi, \K)$ of the CCR algebra $\A_W$ in
which the Hamiltonian $$H = p^2/2 + W(x)$$ is well defined, as a strong limit
of elements of $\A_W$ (on a dense domain), and has a ground state $\Psio \in
\K$.

Moreover, such a representation is independent of $W$, in the class mentioned
above, and it is the  unique non-regular representation  $\pi_0$ in which the
subgroup $V(\b)$, $\b \in \Real$ is regularly represented; its generator $p$ has
a discrete spectrum.

 The Hilbert space $\K$ of $\pi_0$ consists of the formal
sums \be{\psi(x) = \sum_{n \in \Zint} c_n\, e^{i \alpha_n x}, \,\,\,\{c_n\}  \in
l^2(\Cbf), \,\,\, x \in \Real, \,\,\,\a_n \in \Real,}\ee with  scalar product
given by the ergodic mean \be{(\psi, \psi) = \sum_{n \in \Zint} |c_n|^2 =
\lim_{L \ra \infty} (2 L)^{-1} \int_L^L d x\, \bar{\psi}(x)\,\psi(x).}\ee

The Weyl operators are represented by \be{(\pi_o(U(\a)) \psi)(x) = e^{i \a x}
\psi(x), \,\,\,\,\,(\pi_0(V(\b)) \psi)(x) = \psi(x + \b).}\ee

The (orthogonal) decomposition of $\K$ over the spectrum of $V(a)$ is \be{ \K =
\bigoplus_{ \theta \in [0, \,2 \pi)}\H_\theta, \,\,\,\,\,V(a) \,\H_\theta =
e^{i \theta}\,\H_\theta, \,\,\,\,\,\theta \in [0,  2
\pi).}\ee The spectrum of $p$ in $\H_\theta$ is
$\s(p)|_{\H_\theta} = \{ 2 \pi n/a + \theta/a, \,\,\, n \in \Zint\}$ and the
wave functions $\psi_\theta \in \H_\theta$ are quasi periodic of the form
\be{\psi_\theta(x) = e^{i \theta x/a} \sum c_n e^{i 2 \pi n x/a}.}\ee ({\bf
Bloch waves}). The unique ground state is a vector of
$\H_{\theta =0}$.
\end{Proposition}

\sloppy
\section
{Non-regular representations and symmetry breaking} 
\fussy
 We briefly recall that, given a $C^*$-algebra $\A$, an algebraic {\em symmetry}
 is an automorphism 
$\b$ of $\A$; given a state $\om$, the symmetry is {\em
unbroken} in the corresponding representation space if
$\b$ is implemented by a unitary operator $T(\b)$ there, i.e. \be{ \pi_\om(\b(A)) =
T(\b)\,\pi_\om(A) \,T(\b)^*, \,\,\,\,\,\forall A \in \A.}\ee  This means that the
representation
 defined by the state $\om_\b$, $\om_\b(A) \eqq \om(\b(A))$ is unitary equivalent
to  $\pi_\om$: $ \pi_{\om_\b}(A)  = T(\b)\, \pi_\om(A)\, T^*(\b)$ and that  $\omega$ and $\omega_\b$ are described by vectors of the same Hilbert space.

 In this case, $\b$
gives rise to a Wigner symmetry in $\H_\om$, i.e. all
transition amplitudes are invariant. Otherwise, if there is no unitary operator
which implements $\b$ in $\H_\om$, by Wigner theorem on symmetries   at least
one transition amplitude is not invariant and  the symmetry $\b$ is said to be
{\em broken} in $\H_\om$. 

 A one-parameter group $\b^\l$,
 $\l \in \Real$, of symmetries shall be called a {\em continuous symmetry}.
A symmetry is called {\em internal} if it commutes with the one-parameter group $\a_t$, $ t \in \Real$, of the time translations. In the following,   the breaking of an internal symmetry shall be called {\em spontaneous symmetry breaking}.
An algebraic symmetry is said
to be {\em regular} if it maps regular representations
into regular ones. For a discussion of the meaning and the mechanism
of spontaneous symmetry breaking see \cite{FS3}.

In the case of quantum systems described by the canonical Weyl algebra $\A_W$, any
regular algebraic symmetry of $\A_W$ is unbroken in any regular irreducible
representation, since, by Stone-von Neumann theorem, all such representations
are unitarily equivalent. Thus, the important phenomenon of {\em symmetry
breaking}, in the strong sense of a loss of  symmetry as defined above, (which goes much beyond the mere
non-invariance of the ground state) cannot appear in the case of
Heisenberg quantization, more generally in the case of regular Weyl
quantization.

The situation drastically changes in the case of quantum systems whose algebra of observables $\A$ has a non-trivial center. 
A distinguished case is when one has the structure discussed in Section 2, 
namely a canonical algebra $\F_W$ and  an observable (gauge invariant) 
subalgebra $\A$, with a non-trivial center $\Z \subset \A$.

Clearly,  any symmetry $\b$ of $\A$, defined by an element of $\F_W$, is
implemented by a unitary operator $T(\b)$ in the non-regular
representation $\pi$ of $\F_W$, defined by a gauge invariant state $\om_\theta$, 
$ \theta \in \sigma(\Z)$. 
 
However, if $\b$ does not commute with the gauge group $\G$, $\b$   is
broken in {\em each} irreducible representation $\H_\theta$ of the
observable subalgebra $\A$, i.e. $\b$ fails to define a Wigner
symmetry of the gauge invariant states of $\H_\theta = \overline{\A\, 
\Psi_{\omega_\theta}}$, because
$T(\b)$ does not leave $\H_\theta$ invariant. 

 In the regular irreducible representation, $\pi_r$ of $\F_W$, 
the symmetry $\b$ is  unbroken
 but the elements of $\Z$ have a continuous spectrum in
$\H_{\pi_r}$ and there is no gauge invariant (proper)  state vector  in
$\H_{\pi_r}$.\goodbreak

\begin{Proposition} Let $\F_W$ denote the canonical field $C^*$-algebra defined by
a Heisenberg group $G_H$, $\A$ the observable $C^*$-subalgebra, $\Z$ the  
non-trivial 
center of $\A$ generated by the commutative subgroup $\G \subset \G_H$
(gauge group), then \newline i) any algebraic symmetry $\b$ of $\A$, defined by an
element of $G_H$ which does not commute with each element of $\G$, is spontaneously broken in
each irreducible representation of $\A$ ($\theta$ sector);
\newline ii) in any representation of  $\F_W$
defined by a gauge invariant state $\om$, the one-parameter subgroups which do
not commute with $\G$ are non-regularly represented, so that the corresponding
generators cannot be defined as operators in $\H_\om$, only their exponentials
exist.
\end{Proposition}

\Pf\, i) In fact, $\b^\l(\Z) \subseteq \Z$ and since $\Z$ is not left pointwise invariant under $\b^\l$, there is at least one $V \in \Z$, which may be taken unitary, such that $V_\l \eqq \b^\l(V) \neq V$ and in a given  irreducible representation of $\A$,  $V_\l $ and $ V $ are different multiples of the identity. Then, the symmetry breaking condition is realized.    

\noindent ii) Furthermore,  if  $U(\l)$ denotes the one-parameter unitary group which implements $\b^\l$, $ R(\l,V) \eqq V^{-1}\,U(\l)^{-1}\, V \,U(\l) \in \Z$ and is multiple of the identity $e^{i \,\theta(\l, V)} $ in each irreducible representation of $\A$. 
Hence, $\forall A \in \A$, with $< A > \neq 0$, using $V\, U(\lambda)=U(\lambda)\, V\, R(\lambda, V)$, 
$$ < U(\l) \,A > = < V \,U(\l) A\, V^{-1} > = < U(\l)\, A\, R(\l, V) > =$$ $$ e^{i \theta(\l,V)} < U(\l)\,A >,$$
so that 
$$
< U(\l)\,A > =\begin{cases}
0,\text{ if }\l \neq 0,\\ 
< A >, \text{ if }\l = 0,
\end{cases} 
$$
 i.e. $U(\l) $ is not weakly continuous in $\l$.  
\\

For representations of $\A$ defined by a  ground state  $\om_0$,
(more generally by a  state $\om $ invariant under time translations), the 
non-invariance of  $\om_0$, 
\be{\label{Aeqqom}< A > \eqq \om_0(A) \neq \om_0(\b(A)), \,\,\,\,\mbox{for some} \,\,\,\,\, A \in
\A,}\ee  is still compatible with $\b$ giving rise to a Wigner symmetry in the
GNS representation space $\H_{\om_0}$. 
In this case, if $\b$ commutes with the
dynamics, eq.\,\eqref{Aeqqom} implies degeneracy of the ground state. 
This is what
happens if eq.\,\eqref{Aeqqom} holds for $\b$ defined by an element of the field algebra 
$\F_W$ which
commutes also with the gauge group. 


\section{Symmetry breaking by topology and energy spectrum}
\def \limn {\lim_{n \ra \infty}}
 The  spontaneous breaking of  a continuous symmetry in the
quantum theory of infinitely extended systems is usually accompanied by a
strong constraint on the energy spectrum. 

In fact, if the symmetry commutes
with the dynamics (i.e. if the Hamiltonian is symmetric) and it is generated by a conserved current at all times,  the Goldstone theorem
predicts the absence of an energy gap with respect to the ground state, in the
channels related to the ground state by the broken generators and by the operator which provides the order parameter.\footnote{For a
review and critical discussion of the Goldstone theorem see \cite{FS3}.}

The existence of more than one representation for finite-dimensional quantum systems, corresponding to a non-trivial center of the algebra of observables leading to  
 non-regular representations of canonical Weyl field algebra, opens the possibility of spontaneous symmetry breaking also in the case of quantum systems described by  a finite number of canonical variables. In this case, the question arises   
of the implications on the energy spectrum.\goodbreak

 To this purpose, given a $C^*$-algebra $\A$, a one-parameter group $\b^\l$, 
$\l \in \Real$ of automorphisms of $\A$ commuting with the time translations $\a_t$ and a representation $\pi$ of $\A$ 
defined by a ground state $\omega_0$, we consider:

\noindent i) the infinitesimal variation of a generic element $F = \pi(A)$, 
$A \in \A$,
     $$ \delta F =  \delta(\pi(A)) = \frac{d \,\pi(\b^\l(A))}{d \l}|_{\l = 0},$$
ii)  the {\em generation  of the continuous symmetry} $\b^\l$ by elements of the strong closure $\pi(\A)''$ of $\pi(\A)$, 
in the sense that there is a sequence $Q_n = Q_n^* \in \pi(\A)''$, $n=1,..$,  
such that  
\be{ \delta F = i \limn [\,Q_n, \,F\,].}\ee
  If, in the GNS representation defined by $\omega_0$ there is a sequence $Q_n$ which generates $\b^\l$ and converges weakly  to a self adjoint operator $Q$, then $\b^\l$ is 
implementable by the unitary operator $e^{i \l  Q}$, the symmetry is not broken 
and $ < \delta F > \eqq \omega_0(F) \neq 0$ implies that
$\omega_0$ is not invariant, i.e. $Q \Psio \neq 0$. Furthermore, if $\b^\l$ 
commutes  with the time translations $\a_t$, $\Psi_\l \eqq e^{i \l Q} \Psio$  
is a family of {\em degenerate ground states}. 

However, if there is no sequence  $Q_n \in \pi(\A)''$ which generates $\b^\l$ and  
converges weakly to a self-adjoint operator $Q$; then, if $< \delta F > \neq 0$, 
the symmetry is broken  and 
$< \delta F >$ has the meaning of a {\em symmetry breaking order parameter}, 

Furthermore,   by the invariance of the ground state
under $\at$, one has $$ \limn < [ Q_n(t),\,F\,] > = \limn < [\,\a_{t}(Q_n),\, F\,]
> = $$ 
$$= \limn < [\,Q_n,\,\a_{-t}(F)\,] > =  < \delta (\a_{-t}(F)) > = 
 < \delta F > = $$  $$= \limn < [\,Q_n, \,F\, ] >, $$
where the commutation $\b^\l \,\a_t = \a_t \,\b^\l$ has been used in the last but one equality.
\vspace{1mm}

It is worthwhile to stress that such a time independence of the Ward identity 
holds also
in the more general case in which the symmetry does not commute with the
Hamiltonian, but commutes with the time translations in the ground state expectations of the order parameter $F$. For example,  $\limn [\,Q_n(t), \,H\,] = \delta H \neq 0$, but $< [\,\delta H, F\,] > = 0$.  

This is, 
e.g.,  the case in which
the Hamiltonian is invariant up to a time derivative which commutes with $F$ on the ground state
(see the example of the quantum particle on a circle discussed below).

   \def \limn {\lim_{n \ra \infty}}
In conclusion, by the above arguments, for finite dimensional quantum systems one has the following analog of the Goldstone theorem:
\begin{Theorem} (Goldstone) 

Let   
$\A$ be a $C^*$-algebra, $\a_t$ the one-parameter group of automorphisms of $\A$ describing the time traslations   and $\pi$ the  representation of $\A$ defined by  a state
$\omega_0$, invariant under   $\a_t$; if

\noindent i) $\b^\l$, $\l \in \Real$, is a one-parameter group
of automorphisms of the  algebra $\A$,

\noindent ii) there is one  $F \in
\pi(\A)$, such that,  
 $$ < \d \,F > \eqq  d< \b^\l(F)>/d \l|_{\l=0} \neq 0,  $$ 

\noindent iii)
and 
\be{ < \d \,F > =  
i \limn <
[\,Q_n,\,F\,] > = i \limn < [\,Q_n(t),\,F\,] > ,}\label{indipt}\ee  for a suitable sequence of 
$Q_n = Q_n^* \in \pi(\A)''$,  $Q_n(t) \eqq
\at(Q_n)$,  the limit being understood in the sense of convergence of tempered
distributions in the variable $t$, 

\noindent then there is no energy gap above the ground 
state. Actually, there is  a state (Goldstone-like
state) orthogonal to the ground state, with the 
ground
state energy.
\end{Theorem}

\Pf In fact eq.\eqref{indipt} implies that, putting $J_n(t)\equiv 2\ \mathrm{Im} <Q_n U(-t) F>$, 
$$
\lim_{n\to\infty}\tilde J_n(\omega) = <\delta F> \delta(\omega)
$$
therefore the spectral measure of $U(t)$ contains a $\delta(\omega)$.

\vspace{1mm} It is worthwhile to stress that the non-invariance of the ground state expectation
of a field $F$ does not guarantee that one can write a corresponding Ward
identity, eq.\,\eqref{indipt},  a crucial ingredient for the Goldstone theorem.

The interplay between gauge invariance and the breaking of  a
continuous symmetry provides a mechanism for evading the conclusions of the
Goldstone theorem, i.e. for allowing an energy gap in the presence of symmetry
breaking. 

This is typically the case in which the gauge group arises as a consequence of a non-trivial topology ({\bf symmetry breaking by topology}). 

The prototypic realization of such a structure is when the manifold of the configurations of the quantum system has a nontrivial fundamental group leading through its topological invariants (corresponding to its first homology group) to elements of the center of the observable algebra.

The role of the topology in triggering symmetry breaking and affecting its consequences has been realized in specific models of GQFT, through a topological classification of euclidean field configurations. The following theorem provides the general mechanism with no reference to specific (mathematically questionable) ingredients, involving 
the semiclassical approximation.


\begin{Theorem}(Spontaneous symmetry breaking and energy gap)

\noindent Let $\F_W $ be a canonical (field) algebra, $\A$ the observable subalgebra, $\Z$ the center of $\A$ generating gauge transformations on $\F_W$;  if an automorphism $\b$ of $\A$ does not leave  $\Z$ pointwise invariant, in particular if the topological invariants which define elements of $\Z$ are not invariant under $\b$, then $\b$ is spontaneously broken in each irreducible representation of $\A$.
Furthermore, if 

\noindent i) $\b^\l$ is a one-parameter group of  automorphisms of $\A$ defined by elements of $\F_W$,

\noindent ii) in the irreducible representation $\pi_\theta$ of $\A$ defined by a gauge invariant ground state $\omega_\theta$ ($\theta$ vacuum), $\theta \in \sigma(\Z)$,   there is a non-symmetric order parameter
 
\be{ \omega_\theta(\b^\l(A)) \neq \omega_\theta (A), \,\,\,\,\,A \in \A}\ee

\noindent iii) $\b^\l$ commutes with the time translations in the ground state expectations of  the order parameter $A$, i.e 
\be{ \omega_\theta(\b^\l (\a_t(A))) = \omega_\theta(\b^\l(A)),}\ee

\noindent iv) and $\b^\l$ does not leave $\Z$ pointwise invariant, 

\noindent then,  $\b^\l$ cannot be  generated by elements of $\pi(\A)''$, in $\pi_\theta$,  so that the crucial condition of the Goldstone theorem fails and an energy gap is allowed.
\end{Theorem}   

\noindent \Pf The inevitable breaking of $\b$ in any irreducible representation of $\A$ follows as in Proposition 4.1.

\noindent Let us assume that  $\b^\l$ is generated in $\pi_\theta$ by an element of the strong closure of $\pi(\F_W)''$, i.e. there is an operator $Q \in \pi(\F_W)''$ such that  
\be{ \delta A = i \,[\,Q, \,A\,], \,\,\,\,\,\pi_\theta(\delta A) = i\,[\pi_\theta(Q \,A) - \pi_\theta(A\,Q)]\ \,\,\,\,\,\forall A \in \A.}\ee
If $\b^\l$ does not leave  $\Z$ pointwise invariant, there is at least one unitary $V \in \Z$ such that 
$$i  [\,Q, \,V\,] = \delta V \neq 0.$$
Now,  since  $ [ V, \,[\,Q, \,A\,]\,] = 0$ one has
$$   [[\,Q, \,V\,], \,A\,] = -  [[\,A, \,Q\,], \,V\,] -  [\,[\,V,\,A\,], \,Q\,] = 0,$$
i.e.  $ [\,Q, \,V\,] \in \Z$. \goodbreak
 
\noindent  Therefore, by the gauge invariance of $\omega_\theta$,   one has  
$$ \omega_\theta ( Q\,A) = \omega_\theta( V \,Q\,A\,V^{-1}) = \omega_\theta( Q\, V\,A \,V^{-1} +  [\,Q,\,V\,] A V^{-1}) =$$ $$= \omega_\theta(Q\,A) + \omega_\theta([\,Q,\,V\,]\,V^{-1})\,\omega_\theta(A) .$$
Thus, the expectations $\omega_\theta ( Q\,A)$  cannot be defined and one cannot write the symmetry breaking Ward identities which are crucial for the conclusion of the Goldstone theorem. An energy gap is therefore compatible with the spontaneous breaking of $\b^\l$ in irreducible representations of the observable algebra $\A$.  
      
\vspace{2mm} {\bf Remark.} 
In the irreducible regular representation $\pi_r$ of the field algebra $\F_W$,
$\b^\l$ is implemented by a (weakly continuous) group of unitary operators
$T(\l)$, all of the matrix elements are invariant, but there is no gauge invariant state 
(proper) vector  invariant under time translations. The symmetry gets
broken by the direct integral decomposition of $\H_{\pi_r}$ over the spectrum
of $\Z$, but one cannot write a symmetry breaking Ward identity for the
expectation on the gauge invariant ground state.

On the other side, in the representation of $\F_W$ defined by a  gauge
invariant ground state $\om_\theta$, the one-parameter group $T(\l)$ is not
regularly represented. Therefore its generator cannot be defined as an
operator in $\H_{\om_\theta}$ and  $\om_\theta(\d F) \neq 0$
cannot be written in terms of a  commutator. In conclusion, 
the symmetry breaking
Ward identity cannot be written in terms of expectations on $\theta$ states.

\vspace{1mm} \noindent {\bf Examples}. 

\noindent {\bf 1.} {\em Quantum particle on a circle}

\noindent As discussed above, the observable algebra $\A$ is a subalgebra of canonical Weyl (field) algebra $\F_W$ and has a non-trivial center generated by the translations $\gamma^n$ corresponding to the non-trivial fundamental group  of the circle. The one-parameter group of  automorphisms  of $\A$:
\be{ \b^\l(U(n)) = U(n), \,\,\,\,\b^\l(V(\b)) = e^{- i\,\l\,\b}\,V(\b + \l)}\ee 
is realized  by the adjoint action of $U(\l) \in \F_W$: $$\b^\l(U(n)\, V(\b)) = U(\l) \,U(n)\,V(\b)\, U(-\l).
$$
Actually, $\F_W$ is the minimal extension of $\A$ such that the automorphisms $\b^\l$ are inner and $\F_W$ includes the operators which intertwine between inequivalent representations of $\A$. This justifies the use of the (canonical) field algebra $\F_W$ for a canonical description of both the observables and {\em all} the states of the quantum system.

In the representation $\pi_\theta$ of $\A$ there is a non symmetric order parameter:
$$ \omega_\theta(\b^\l(V(\b))) = e^{- i\,\l\,\b} \,\omega_\theta(V(\b)), \,\,\,\,\omega_\theta(V(\b)) = e^{i \,\theta\,\b/2\pi},$$ and $\b^\l$ commutes with the time translations $e^{ i \,H t}\,\,H = p^2/2 m$ in expectations of the order parameter: $\omega_\theta(\b^\l\,\a_t(V(\b))) = \omega_\theta(\a_t \b^\l(V(\b))).$ 

Furthermore, $\b^\l$ does not leave the center of $\A$ pointwise invariant. Theorem 5.2 applies and an energy gap is allowed; in fact, in the representation $\pi_\theta$ of $\A$ the spectrum of the Hamiltonian is discrete,  given by  
$$ \sigma^\theta(H) = \frac{(n + \theta/2\pi)^2}{2m}, \,\,\,\,\,n \in \Zint,$$
 and displays an energy  gap above the ground state energy $(\theta/2\pi)^2/2m$, compatibly with the spontaneous symmetry breaking of $\b^\l$. It is worthwhile to note that the energy gap is provided by the spectrum of the center, i.e. by the (discrete) spectrum of the fundamental group.  

\vspace{2mm}

\noindent {\bf 2.} {\em Bloch electron}

\noindent For simplicity, we consider the one-dimensional case. The field algebra $\F_W$ is the canonical Weyl algebra and the observable subalgebra is characterized by its pointwise invariance under the gauge transformations defined by the adjoint action of the center of $\A$, which arises from the non-trivial topology of $\Real/[0,\,a]$. The automorphisms $\b^\l$ defined  by the Weyl operators $U(\l)$, do not leave the center pointwise invariant. In the representation $\pi_\theta$, defined by the gauge invariant  states $\omega_\theta$, there is a symmetry breaking order parameter $\omega_\theta(V(\b))$. 

Furthermore, $\b^\l$ commutes with $T(\t) \eqq e^{-i \t\,p^2/2m }$ on expectations of the order parameter
$$ \omega_\theta( \b^\l (T(\t)\,V(\b)\,T(\t)^{-1}))= \omega_\theta( \b^\l (V(\b))) =$$ $$= \omega_\theta((T(\t)\,\b^\l(V(\b))\,T(\t)^{-1})$$ since the  states $\omega_\theta$ are invariant under by the one-parameter group $T(\t)$.  Thus, if $\b^\l$ would be generated by $q$ in the representation $\pi_\theta$ one could apply the Goldstone theorem and  conclude that the spectrum of $p^2$ does not have a gap above the ground state value. However,   Theorem 5.2 applies and in fact the spectrum of $p^2$  has a gap 
given by the spectrum  of the fundamental group of $\Real/[0, \,a]$.


\begin{thebibliography}{9}

\bibitem
{FS1} F. Strocchi, {\em An Introduction to the Mathematical Structure of Quantum Mechanics}, 2nd expanded edition World Scientific 2010

\bibitem
{FS2} F. Strocchi, The physical principles of quantum mechanics. A critical review, Eur. Phys. J. Plus (2012), {\bf 127}:12 

\bibitem
{AMS1} F. Acerbi, G. Morchio and F. Strocchi, Jour. Math. Phys. {\bf 34}, 899 (1992) 

\bibitem
{AMS2} F. Acerbi, G. Morchio and F. Strocchi, Lett. Math. Phys. {\bf 26}, 13 (1992)

\bibitem
{AMS3} F. Acerbi, G. Morchio and F. Strocchi, Lett. Math. Phys. {\bf 27}, 1 (1993) 


\bibitem
{FSS} F. Strocchi, {\em Gauge invariance and Weyl-polymer quantization}, to be published

\bibitem
{MS}  \G. Morchio and F. Strocchi, Lett. Math. Phys. {\bf 82}, 219 (2007) and work in preparation 

\bibitem
{Stone}
M. H. Stone, Proc. Natl. Acad. of Sci. U.S.A.  {\bf 16}, 172 (1930)

\bibitem
{vonNeumann}
J. von Neumann, Math. Ann. {\bf104}, 570 (1931)

\bibitem
{Mackey}
G. Mackey,
Duke Math. J.
{\bf16}, 313 (1949)

\bibitem{Rieffel}
M. A. Rieffel,
Duke Math. J.
{\bf 39}, 745 (1972)

\bibitem 
{ZakOr} J. Zak, Phys. Rev. {\bf 168}, 686 (1968) 

\bibitem
{CMS} S. Cavallaro, G. Morchio and F. Strocchi, Lett. Math. Phys. {\bf 47}, 307 (1999)

\bibitem
{BMPS} R. Baume, J. Manuceau, A Pellet and M. Siruge, Comm. Math. Phys. {\bf 38}, 29 (1974)

\bibitem
{J} R. Jackiw, 
Topological Investigations of Quantized Gauge Theories, in S.B. Treiman, 
R. Jackiw, B. Zumino and E. Witten, {\em Current Algebra and Anomalies}, 
World Scientific 1985
 

\bibitem
{LMS} J. L\"{o}ffelholz, G. Morchio and F. Strocchi, Lett. Math. Phys. {\bf 35}, 251 (1995)




\bibitem
{FS3} F. Strocchi, {\em
Symmetry Breaking}, 2nd ed.,  Springer 2008, Chap. 15





\end{thebibliography}
\end{document}